# Circuits-Informed Machine Learning Technique for Blind Open-Loop Digital Calibration of SAR ADC

Sumukh Bhanushali†, Debnath Maiti†, Phaneendra Bikkina‡, Esko Mikkola‡, Arindam Sanyal†

† Arizona State University, ‡ Alphacore, Inc.

SAR architecture is widely adopted due to its high energy efficiency at medium resolution. At >10-bit resolution and sampling frequency >100MHz, performance of SAR ADC is limited by sampling noise, comparator noise and offset, and reference ripple. Several techniques exist to suppress these non-idealities, such as stochastic estimation for reducing comparator noise [1], kT/C noise cancellation [2], comparator offset calibration in loop-unrolled SAR [3], and reference ripple suppression [4-7]. These techniques are algorithm-based which target specific errors and require knowledge of the errors to correct them. In addition, some of these techniques need analog solutions such as low-noise amplifiers for kT/C noise cancellation and low-output resistance buffers for reference regulation. In contrast, this work proposes a machine-learning (ML) approach for blind digital calibration that simultaneously suppresses multiple errors without requiring prior knowledge of these errors. The key advantage of the proposed technique is that the ML model can correct even errors unforeseen during design and are generalizable to multiple architectures unlike algorithm-based techniques. In the proposed technique, a supervised ML model uses the difference between the main SAR ADC and a low-speed reference SAR ADC to learn the errors in the main ADC and suppress them. A few recent works have applied ML to improve ADC performance [8-10]. Ref. [8] applies ML to correct mismatches in time-interleaved ADC, but only achieves a modest improvement in ENOB of ~1.5-bits even when using a complex convolutional neural network with >13,000 parameters. Ref. [9] employs unsupervised ML on a split-ADC architecture to correct errors by forcing the two ADCs to match each other. However, the unsupervised approach may not always converge to the desired operating point and requires each ADC to follow a different conversion trajectory that is neither trivial to realize nor easily generalizable to different architectures. Ref. [10] demonstrates supervised ML calibration on a 1MHz SAR ADC that does not capture errors introduced at high-speed, e.g. reference ripple. In addition, the reference ADC model is extracted in software from the main ADC which limits applicability of the calibration technique. In contrast, the proposed ML technique requires 9x less parameters than the ML model in [8], will always converge due to supervised ML training, implements both reference and main ADCs in silicon and calibrates both static and dynamic errors introduced at high-speed.

The proposed architecture is shown in Fig.1. The main ADC is a 12-bit, top-plate sampled synchronous ADC with 1-bit redundancy and unit capacitor of 0.6fF. The ADC uses bi-directional single-sided switching (BDSS) technique to reduce switching energy [10]. In this work, the reference ADC used for supervised ML training is a copy of the main ADC but runs at Fs/8 where the Fs is the sampling speed of the main ADC. The two ADCs do not need to be matched accurately since the ML model calibrates mismatches between the two ADCs as well. The sampling instants of the main and reference ADCs align every 8 samples of the main ADC, and the ML model is trained during these aligned sampling instants. To reduce the training requirement, the ML model is trained to estimate only the errors between the main and reference ADC outputs rather than letting the ML model learn the main ADC transfer function. While the training happens only when the sampling instants of the main and reference ADCs are aligned, the trained ML model performs calibration for all the main ADC outputs. In-order to achieve high SFDR without requiring a complex deep neural network for calibration, this work proposes a circuits-informed approach that allows the use of a shallow artificial neural network (ANN). The proposed technique derives insights from ADC behavior to extract key features for training the ANN as shown in Fig. 1. Static errors, such as capacitance mismatch and comparator offset can be estimated by comparing the main ADC and reference ADC outputs directly when they align. On the other hand, dynamic errors, such as reference ripple, have memory effect, and affect multiple ADC outputs. Hence, the current and several past samples of the main ADC output are used as features to the ML model for calibrating both dynamic and static errors. An additional non-ideality is introduced in the proposed architecture since the sampling instants of the main and reference ADCs cannot be precisely matched even when they align. This introduces relative gain and timing mismatches between the reference and main ADCs. To account for this sampling instant mismatch, temporal derivative of the main ADC output is used as an additional feature that captures both gain and timing mismatch with reference ADC. While the reference ADC used in this work is a copy of the main ADC, its dynamic errors are significantly suppressed since it runs at 8x lower speed. However, the reference ADC still needs calibration for static capacitor mismatch. A look-up table (LUT) based approach is used for foreground calibration of the reference ADC. As shown in Fig. 2, a low frequency sinusoidal input is applied to the reference ADC and its histogram is used to reconstruct ADC transfer function. Due to capacitor mismatches, the transition points in the transfer function are unevenly spaced. A straight line is fitted through the end-points of the ADC transfer function to derive ideal ADC outputs. The accuracy of this ideal transfer function is limited by noise in the ADC and hence an averaged transfer function is derived by repeating this process 100 times. An LUT is created in which each ADC code is mapped to the ideal ADC code obtained from the averaged transfer function. As shown in Fig. 2, this LUT-based technique improves SNDR by >3dB and SFDR by >15dB.

A test-chip fabricated in 28nm CMOS (Fig. 7) consumes 1.2mW from 0.9V supply with the main ADC consuming 1.1mW at Fs=84MHz and the reference ADC consuming 0.1mW at Fs=10.5MHz, thus resulting in only 10% power overhead. The ML model is synthesized digitally and is estimated to consume 0.91mW from 0.5V supply (0.62mW for training and 0.29mW for inference) with 12-bit fixed-point precision. Fig. 3 plots the measured FFT of the main ADC before and after ML correction for low frequency and near-nyquist input. The uncorrected ADC performance is limited primarily by reference ripple which is exacerbated since no reference decoupling capacitor or on-chip reference buffers have been used in this work. ML calibration improves SNDR/SFDR by 32/44.8dB and by 30.5/38.6dB at low and near-nyquist input frequencies respectively for 1V pk-pk amplitude. The ANN used for calibration has one hidden layer with tanh activation and an output layer with one neuron and linear activation. Fig. 4 plots the improvements in SNDR and SFDR as a function of the number of hidden neurons and number of features ($m$). The ANN performance initially improves and then saturates as the model size becomes larger. For this work, the number of hidden neurons and number of features are selected as 40 and 41 respectively which corresponds to 1680 trainable parameters. The small size of the ANN model is a key enabler for it to be included in the back-end DSP for calibration. The ML training converges after $2^{16}$ samples and improves DNL/INL from 212.6/-120.2 LSB to 0.8/1.4 LSB respectively. Fig. 5 plots SNDR and SFDR before and after ML correction as a function of input frequency. ML calibration improves SNDR/SFDR by >28/35dB over the entire signal band. A two-tone test is performed with input tones at 1.03MHz and 1.09MHz. Memory effect due to reference ripple creates higher-order nonlinearities and results in multiple spurs spaced by the difference between the input tones. ML correction improves IM2/IM3 by 30.8/38dB respectively. Fig. 6 compares the proposed ML-based calibration technique with other calibration techniques for SAR ADCs. Compared to the algorithm-based techniques that target individual errors, ML techniques can calibrate all errors using a single neural network. Compared to the ML technique in [9], the proposed method has 80x faster convergence time with 3x less neurons thanks to curated feature extraction which incorporates ADC knowledge into the ML training. Power consumption of the ML model can be further reduced through in-memory computing (IMC) techniques [12] and using analog activation functions [13]. Since the ML output captures ADC error that has small dynamic range (~7-8 bits), the final stage readout ADC is not expected to consume significant power if IMC is adopted. This work has demonstrated the potential advantage of supervised ML over algorithm-based techniques for calibration of ADCs in advanced nodes without requiring prior knowledge of non-idealities.

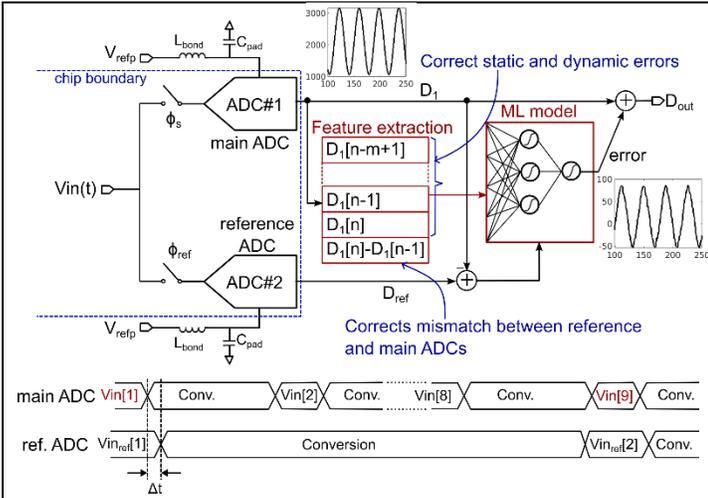

Fig. 1. Schematic and timing diagram of the proposed ADC calibration technique using supervised ML technique and low-speed reference ADC

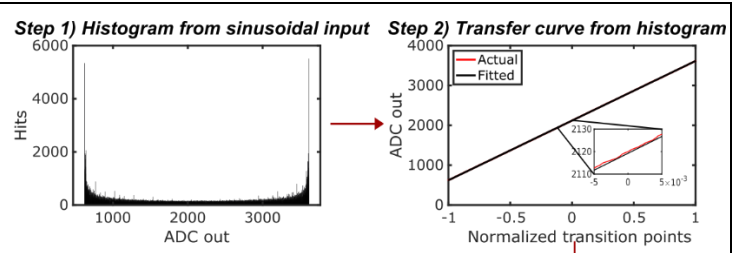

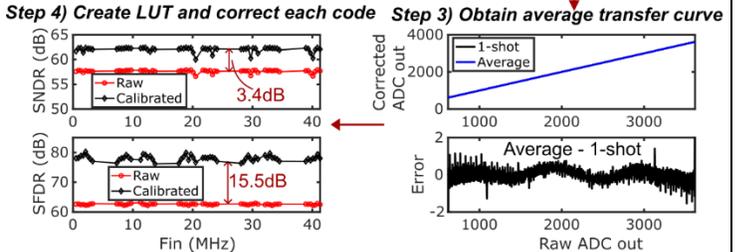

Fig. 2. The foreground calibration technique used on reference ADC that improves SNDR by >3dB and SFDR by >15dB

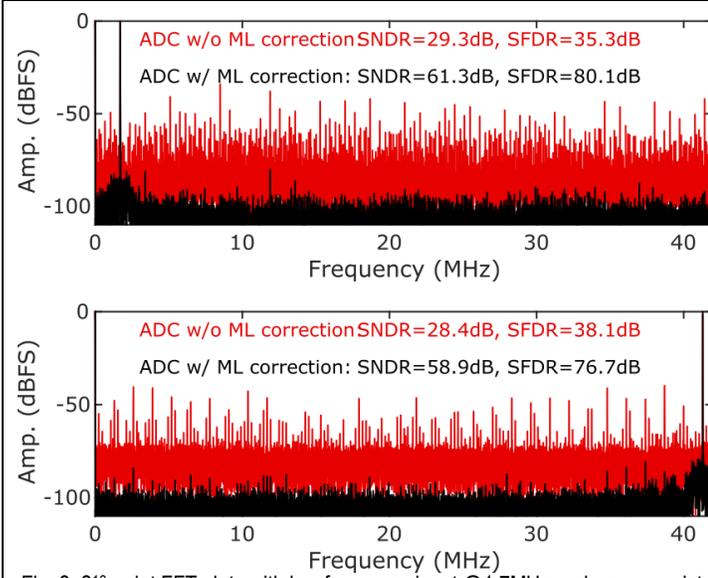

Fig. 3. $2^{16}$-point FFT plots with low frequency input @1.7MHz and near-nyquist input at @41.3MHz before and after ML based calibration

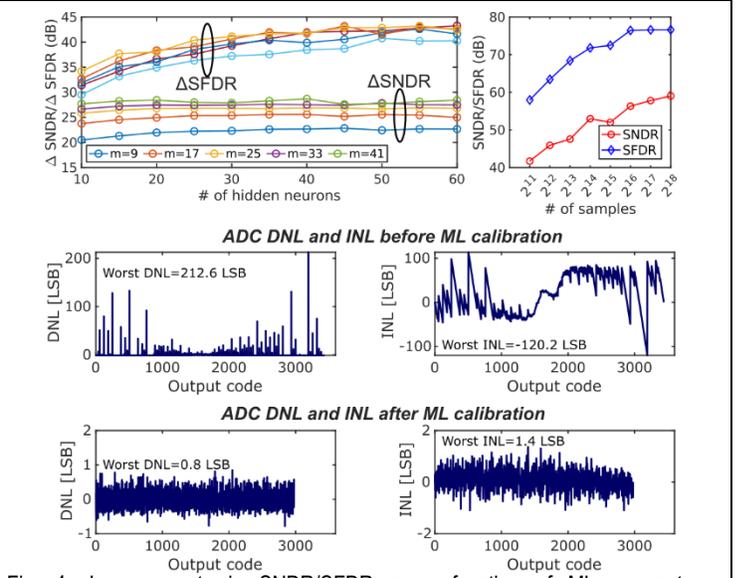

Fig. 4. Improvements in SNDR/SFDR as a function of ML parameters, convergence performance and DNL and INL before and after ML calibration

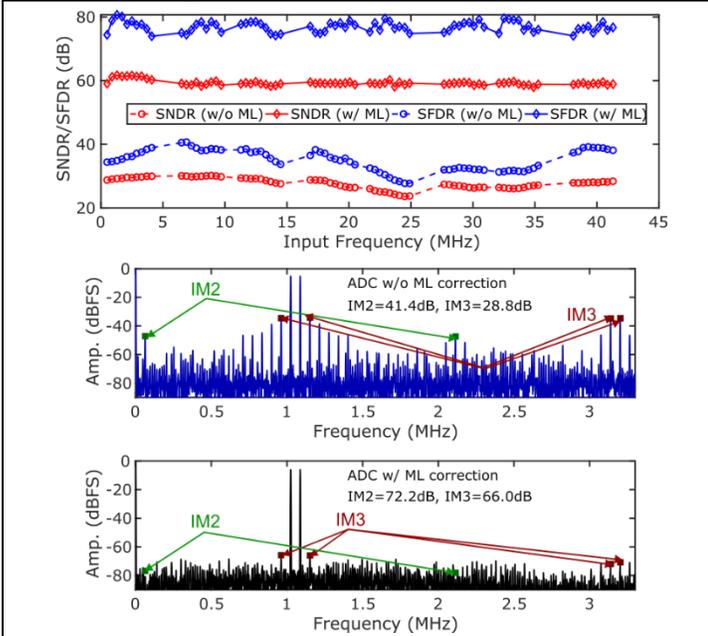

Fig. 5. SNDR/SFDR as a function of input frequency and results of two-tone test before and after ML based calibration

|  | [1] | [2] | [4] | [3] | [9] | This Work |
|---|---|---|---|---|---|---|
| Process (nm) | 65 | 40 | 40 | 40 | 28 | 28 |
| Resolution (bit) | 11 | 13 | 10 | 6-8 | 14 | 12 |
| Supply (V) | 0.7 | 0.8 | 1.2 | 2.5/1.1 | 1 | 0.9 |
| Samp. cap (pF) | 1 | 0.24/0.3 | 1 | - | - | 2.5 |
| Fs (MHz) | 0.1 | 40 | 120 | 900 | 60 | 84 |
| Correction technique | Bayesian estimation | CDS[1] | Replica DACs | Comp. Offset | Machine learning | |
| Error corrected | Quant. err, Comp. noise | kT/C noise | Ref. ripple | Offset | All non-linearities/errors | |
| Correction type | Digital | Analog | Analog | Digital | | |
| Conv. time (samples) | N/A | | | N/A | 30M@ 60MHz | 64k@ 10.5MHz |
| # of ML neurons | N/A | | | N/A | 129 | 41 |
| SNDR w/o cal (dB) | 59.4 | - | 47.4 | - | 37.3 | 29.3 |
| SNDR w/ cal (dB) | 64.5 | 69 | 57.5 | 33.4 | 63.9 | 61.3 |
| SFDR w/o cal (dB) | - | - | 62.7 | - | 63.4 | 35.3 |
| SFDR w/ cal (dB) | 78 | 79.2 | 71.7 | 48.4 | 83.3 | 80.1 |
| Power (μW) | 0.6 | 591 | 1120 | 7000 | 2790[2] | 1200[3]/2110[4] |
| FoM$_w$ (fJ/step) | 4.5 | 6.4 | 20.5 | 20[2] | 28 | 14.7[3]/25.8[4] |

[1]CDS: correlated double sampling; [2]only ADC; [3]Dual ADC; [4]Dual ADC+ML

Fig. 6. Comparison with state-of-the-art SAR ADCs with error correction techniques

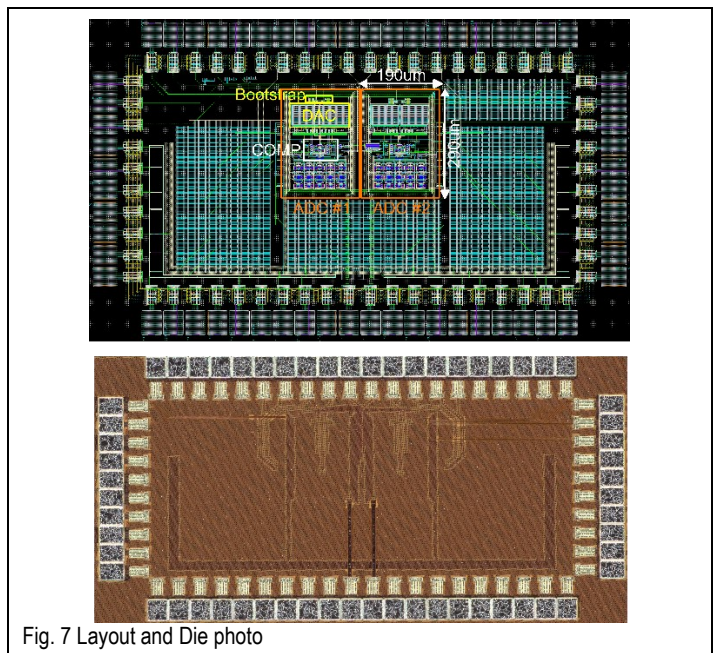

Fig. 7 Layout and Die photo